# Spin-to-orbital angular momentum conversion in focusing, scattering, and imaging systems

Konstantin Y. Bliokh[1,2,*], Elena A. Ostrovskaya[3], Miguel A. Alonso[4], Oscar G. Rodríguez-Herrera[1], David Lara[5], and Chris Dainty[1]

[1] Applied Optics Group, School of Physics, National University of Ireland, Galway, Galway, Ireland
[2] A. Usikov Institute of Radiophysics and Electronics, 12 Ak. Proskury St., Kharkov 61085, Ukraine
[3] Nonlinear Physics Centre, Research School of Physics and Engineering, The Australian National University, Canberra ACT 0200, Australia
[4] The Institute of Optics, University of Rochester, Rochester, New York, 14627, USA
[5] The Blackett Laboratory, Imperial College London, SW7 2BW, London, United Kingdom
*k.bliokh@gmail.com

**Abstract:** We present a general theory of spin-to-orbital angular momentum (AM) conversion of light in focusing, scattering, and imaging optical systems. Our theory employs universal geometric transformations of non-paraxial optical fields in such systems and allows for direct calculation and comparison of the AM conversion efficiency in different physical settings. Observations of the AM conversions using local intensity distributions and far-field polarimetric measurements are discussed.

**References and links**

1. L. Allen, S. M. Barnett, and M. J. Padgett, editors, *Optical angular momentum* (Taylor and Francis, 2003).
2. S. Franke-Arnold, L. Allen, and M. Padgett, "Advances in optical angular momentum," Laser & Photon. Rev. **2**, 299-313 (2008).
3. G. Biener, A. Niv, V. Kleiner, and E. Hasman, "Formation of helical beams by use of Pancharatnam-Berry phase optical elements," Opt. Lett. **27**, 1875-1877 (2002).
4. A. Ciattoni, G. Cincotti, and C. Palma, "Angular momentum dynamics of a paraxial beam in a uniaxial crystal," Phys. Rev. E **67**, 036618 (2003).
5. L. Marrucci, C. Manzo, and D. Paparo, "Optical spin-to-orbital angular momentum conversion in inhomogeneous anisotropic media," Phys. Rev. Lett. **96**, 163905 (2006).
6. G. F. Calvo and A. Picón, "Spin-induced angular momentum switching," Opt. Lett. **32**, 838-840 (2007).
7. E. Brasselet, Y. Izdebskaya, V. Shvedov, A. Desyatnikov, W. Krolikowski, and Y. S. Kivshar, "Dynamics of optical spin-orbit coupling in uniaxial crystals," Opt. Lett. **34**, 1021-1023 (2009).
8. M. Y. Darsht, B. Y. Zel'dovich, I. V. Kataevskaya, and N. D. Kundikova, "Formation of an isolated wavefront dislocation," Zh. Eksp. Teor. Fiz. **107**, 1464 **[**JETP **80**, 817 (1995)**]**.
9. Z. Bomzon, M. Gu, and J. Shamir, "Angular momentum and geometric phases in tightly-focused circularly polarized plane waves," Appl. Phys. Lett. **89**, 241104 (2006).
10. Z. Bomzon and M. Gu, "Space-variant geometrical phases in focused cylindrical light beams," Opt. Lett. **32**, 3017-3019 (2007).
11. Y. Zhao, J. S. Edgar, G. D. M. Jeffries, D. McGloin, and D. T. Chiu, "Spin-to-orbital angular momentum conversion in a strongly focused optical beam," Phys. Rev. Lett. **99**, 073901 (2007).
12. Y. Zhao, D. Shapiro, D. McGloin, D. T. Chiu, and S. Marchesini, "Direct observation of the transfer of orbital angular momentum to metal particles from a focused circularly polarized Gaussian beam," Opt. Express **17**, 23316-23322 (2009).
13. T. A. Nieminen, A. B. Stilgoe, N. R. Heckenberg, and H. Rubinsztein-Dunlop, "Angular momentum of a strongly focused Gaussian beam," J. Opt. A: Pure Appl. Opt. **10**, 115005 (2008).
14. Y. Gorodetski, A. Niv, V. Kleiner, and E. Hasman, "Observation of the spin-based plasmonic effect in nanoscale structures," Phys. Rev. Lett. **101**, 043903 (2008).
15. P. B. Monteiro, P. A. M. Neto, and H. M. Nussenzveig, "Angular momentum of focused beams: Beyond the paraxial approximation," Phys. Rev. A **79**, 033830 (2009).
16. O. G. Rodríguez-Herrera, D. Lara, K. Y. Bliokh, E. A. Ostrovskaya, and C. Dainty, "Optical nanoprobing via spin-orbit interaction of light," Phys. Rev. Lett. **104**, 253601 (2010).

17. M. R. Foreman and P. Török, "Spin-orbit coupling and conservation of angular momentum flux in non-paraxial imaging of forbidden radiation," New J. Phys. **13**, 063041 (2011).
18. C. Schwartz and A. Dogariu, "Conservation of angular momentum of light in single scattering," Opt. Express **14**, 8425-8433 (2006).
19. C. Schwartz and A. Dogariu, "Backscattered polarization patterns, optical vortices, and the angular momentum of light," Opt. Lett. **31**, 1121-1123 (2006).
20. C. Schwartz and A. Dogariu, "Backscattered polarization patterns determined by conservation of angular momentum," J. Opt. Soc. Am. A **25**, 431-436 (2008).
21. D. Haefner, S. Sukhov, and A. Dogariu, "Spin Hall Effect of light in spherical geometry," Phys. Rev. Lett. **102**, 123903 (2009).
22. Y. Gorodetski, N. Shitrit, I. Bretner, V. Kleiner, and E. Hasman, "Observation of optical spin symmetry breaking in nanoapertures," Nano Lett. **9**, 3016-3019 (2009).
23. L. T. Vuong, A. J. L. Adam, J. M. Brok, P. C. M. Planken, and H. P. Urbach, "Electromagnetic spin-orbit interactions via scattering of subwavelength apertures," Phys. Rev. Lett., **104**, 083903 (2010).
24. Y. Gorodetski, S. Nechaev, V. Kleiner, and E. Hasman, "Plasmonic Aharonov-Bohm effect: Optical spin as the magnetic flux parameter," Phys. Rev. B **82**, 125433 (2010).
25. E. Hasman, G. Biener, A. Niv, and V. Kleiner, "Space-variant polarization manipulation," Prog. Opt. **47**, 215-289 (2005).
26. L. Marrucci, E. Karimi, S. Slussarenko, B. Piccirillo, E. Santamato, E. Nagali, and F. Sciarrino, "Spin-to-orbital conversion of the angular momentum of light and its classical and quantum applications," J. Opt. **13**, 064001 (2011).
27. S. M. Barnett and L. Allen, "Orbital angular-momentum and nonparaxial light-beams," Opt. Commun. **110**, 670-678 (1994).
28. C.-F. Li, "Spin and orbital angular momentum of a class of nonparaxial light beams having a globally defined polarization," Phys. Rev. A **80**, 063814 (2009).
29. K. Y. Bliokh, M. A. Alonso, E. A. Ostrovskaya, and A. Aiello, "Angular momenta and spin-orbit interaction of nonparaxial light in free space," Phys. Rev A **82**, 063825 (2010).
30. N. B. Baranova, A. Y. Savchenko, and B. Y. Zel'dovich, "Transverse shift of a focal spot due to switching of the sign of circular-polarization," JETP Lett. **59**, 232-234 (1994).
31. B. Y. Zel'dovich, N. D. Kundikova, and L. F. Rogacheva, "Observed transverse shift of a focal spot upon a change in the sign of circular polarization," JETP Lett. **59**, 766-769 (1994).
32. V. Garbin, G. Volpe, E. Ferrari, M. Versluis, D. Cojoc, and D. Petrov, "Mie scattering distinguishes the topological charge of an optical vortex: a homage to Gustav Mie," New J. Phys. **11**, 013046 (2009).
33. E. Brasselet, N. Murazawa, H. Misawa, and S. Juodkazis, "Optical vortices from liquid crystal droplets," Phys. Rev. Lett. **103**, 103903 (2009).
34. F. Manni *et al.*, "Spin-to-orbital angular momentum conversion in semiconductor microcavities," Phys. Rev. B **83**, 241307(R) (2011).
35. S. J. van Enk and G. Nienhuis, "Spin and orbital angular momentum of photons," Europhys. Lett. **25**, 497-501 (1994).
36. S.J. van Enk and G. Nienhuis, "Commutation rules and eigenvalues of spin and orbital angular momentum of radiation fields," J. Mod. Opt. **41**, 963-977 (1994).
37. E. Wolf, "Electromagnetic diffraction in optical systems. I. An integral representation of the image field," Proc. Roy. Soc. London. Ser. A **253**, 349-357 (1959).
38. B. Richards and E. Wolf, "Electromagnetic diffraction in optical systems. II. Structure of the image field in an aplanatic system," Proc. Roy. Soc. London. Ser. A **253**, 358-379 (1959).
39. M. V. Berry, "Interpreting the anholonomy of coiled light," Nature **326**, 277-278 (1987).
40. P. Török, P. D. Higdon, and T. Wilson, "On the general properties of polarized light conventional and confocal microscopes," Opt. Commun. **148**, 300-315 (1998).
41. Bekshaev, A., Bliokh, K. Y. and Soskin, M. (2011). Internal flows and energy circulation in light beams. *J. Opt.*, **13**, 053001.
42. R. Bhandari, "Polarization of light and topological phases," Phys. Rep. **281**, 2-64 (1997).
43. M. A. Alonso and G. W. Forbes, "Uncertainty products for nonparaxial wave fields," J. Opt. Soc. Am. A **17**, 2391-2402 (2000).
44. M. A. Alonso, "The effect of orbital angular momentum and helicity in the uncertainty-type relations between focal spot size and angular spread," J. Opt. **13**, 064016 (2011).
45. N. Bokor, Y. Iketaki, T. Watanabe, and M. Fujii, "Investigation of polarization effects for high-numerical-aperture first-order Laguerre-Gaussian beams by 2D scanning with a single fluorescent microbead," Opt. Express **13**, 10440-10447 (2005).
46. Y. Iketaki, T. Watanabe, N. Bokor, and M. Fujii, "Investigation of the center intensity of first- and second-order Laguerre-Gaussian beams with linear and circular polarization," Opt. Lett. **32**, 2357-2359 (2007).
47. K. Y. Bliokh, Y. Gorodetski, V. Kleiner, and E. Hasman, "Coriolis effect in optics: Unified geometric phase and spin-Hall effect," Phys. Rev. Lett. **101**, 030404 (2008).
48. M. Born and E. Wolf, *Principles of Optics*, 7th edn. (Pergamon, 2005).
49. G. Moe and W. Happer, "Conservation of angular momentum for light propagating in a transparent anisotropic medium," J. Phys. B: Atom. Molec. Phys. **10**, 1191-1208 (1977).

50. V. Rossetto and A. C. Maggs, "Writhing geometry of stiff polymers and scattered light," Eur. Phys. J. B **29**, 323-326 (2002).
51. D. Lacoste, V. Rossetto, F. Jaillon, and H. Saint-Jalmes, "Geometric depolarization in patterns formed by backscattered light," Opt. Lett. **29**, 2040-2042 (2004).
52. M. V. Berry, "Optical currents," J. Opt. A: Pure Appl. Opt. **11**, 094001 (2009).
53. A. Bekshaev and S. Sviridova, "Mechanical action of inhomogeneously polarized optical fields and detection of the internal energy flows," arXiv:1102.3514.
54. O. G. Rodríguez-Herrera, D. Lara, and C. Dainty, "Far-field polarization-based sensitivity to ssubresolution displacements of a sub-resolution scatterer in tightly focused fields," *Opt. Express*, **18**, 5609-5628 (2010).
55. Y.A. Kravtsov, B. Bieg, and K.Y. Bliokh, "Stokes-vector evolution in a weakly anisotropic inhomogeneous medium," J. Opt. Soc. Am. A **24**, 3388-3396 (2007).
56. M. V. Berry, "Paraxial beams of spinning light," Proc. SPIE **3487**, 6-11 (1997).

## 1. Introduction

Together with energy and momentum, angular momentum (AM) is one of the most important dynamical characteristics of light [1,2]. For paraxial fields in free space, the eigenmodes of the AM operator are circularly-polarized vortex beams, where polarization helicity $\sigma = \pm 1$ specifies the value of spin AM (SAM) per photon, whereas the vortex charge $\ell = 0, \pm 1, \pm 2, ...$ yields the orbital AM (OAM) per photon. Generation of the helicity-dependent vortices in optical systems signifies the spin-to-orbital AM conversion. This phenomenon is attracting noticeable attention in recent years and occurs in two basic situations: (A) upon interaction of paraxial light with anisotropic media possessing certain azimuthal symmetries [3-8] and (B) in essentially non-paraxial optical fields in locally-isotropic media [8-24]. (Optical fibers [8] can be regarded as an intermediate case.) In the case (B), nonparaxial AM states of light appear mostly upon (i) tight focusing [9-17], (ii) scattering by small particles or apertures [16-24], and (iii) in high numerical aperture (NA) imaging of small particles [16,17]. Importantly, the case (iii), involving a combination of focusing and scattering processes, represents a fundamental mechanism for translating the fine spin-orbit effects at micro- and nano-scales to the far-field.

The spin-to-orbital AM conversion in anisotropic paraxial systems is an *extrinsic* phenomenon produced by the azimuthally-dependent phase difference between ordinary and extraordinary modes, which is well-studied and reviewed [25,26]. On the other hand, the AM conversion in nonparaxial fields owes its origin to the intrinsic properties of light, geometric Berry phases, and fundamental separation of the SAM and OAM in the generic nonparaxial case [27-29]. The spin-to-orbital conversion in nonparaxial light has been considered for different systems using different *ad hoc* methods, such as Debye-Wolf theory for focusing or Mie theory for scattering. Furthermore, the spin-orbit interaction in a variety of similar imaging schemes is ascribed either to focusing [30,31], or to scattering [32], or to anisotropy [33,34]. Obviously, all these mechanisms co-exist and their unifying description and discrimination is necessary.

In the present paper we examine the spin-to-orbital AM conversion that appears in nonparaxial optical fields interacting with locally-isotropic media upon (i) focusing, (ii) scattering, and (iii) imaging. We develop a unifying theory of these effects based on universal, purely geometrical transformations of the fields and fundamental AM operators. Our approach highlights the common geometric origin of AM conversion due to different optical processes and allows us to compare the conversion efficiency in different physical settings depending on the aperture angles and properties of the incoming light.

## 2. Basic equations

We consider monochromatic wave electric fields $\mathcal{E}(\mathbf{r},t)$ characterized by their complex amplitudes $\mathbf{E}(\mathbf{r})$: $\mathcal{E}(\mathbf{r},t) = \text{Re}\left[\mathbf{E}(\mathbf{r})e^{-i\omega t}\right]$. Similar relations are implied when the wave magnetic field $\mathbf{H}(\mathbf{r})$ is involved. The Fourier or angular spectra of the fields are denoted by $\tilde{\mathbf{E}}(\mathbf{k})$ and $\tilde{\mathbf{H}}(\mathbf{k})$. Throughout the paper we consider optical systems with axial symmetry about the $z$-axis, and, for the AM description, it is convenient to use the basis of circular polarizations with respect to the optical axis. Given that $(\mathbf{u}_x, \mathbf{u}_y, \mathbf{u}_z)$ are the basic vectors of the laboratory Cartesian frame and $\mathbf{E}_L = (E_x, E_y, E_z)^T$ is the vector of complex amplitudes of a monochromatic electric field in this basis, the basic vectors and electric field components in the circular basis are:

$$\mathbf{u}^\pm = \frac{\mathbf{u}_x \pm i\mathbf{u}_y}{\sqrt{2}}, \quad E^\pm = \frac{E_x \mp iE_y}{\sqrt{2}}, \tag{1}$$

so that $\mathbf{E}_C = (E^+, E^-, E_z)^T$. Thus, transition from the Cartesian to circular basis is realized via the unitary transformation

$$\mathbf{E}_L = \hat{V} \mathbf{E}_C, \quad \hat{V} = \frac{1}{\sqrt{2}} \begin{pmatrix} 1 & 1 & 0 \\ i & -i & 0 \\ 0 & 0 & \sqrt{2} \end{pmatrix}. \tag{2}$$

Throughout the paper we mostly analyze the fields in the circular basis, and the subscripts $C$ are omitted.

The operators of the $z$-component of the OAM and SAM of light in the Cartesian basis are [1]: $\hat{L}_z = -i\partial/\partial\phi$ and $(\hat{S}_z)_{ij} = -i\varepsilon_{zij}$, where $\phi$ is the azimuthal angle (either in coordinate or momentum space, depending on representation) and $\varepsilon_{ijl}$ is the Levi-Civita symbol. In the circular basis, these operator become, correspondingly

$$\hat{L}_z = -i\frac{\partial}{\partial\phi}, \quad \hat{V}^\dagger \hat{S}_z \hat{V} = \hat{\sigma} = \mathrm{diag}(1, -1, 0). \tag{3}$$

Note that the vortex fields $\mathbf{E} \propto \exp(i\ell\phi)$ are eigenmodes of $\hat{L}_z$ with the eigenvalues $\ell$. At the same time, the circularly polarized paraxial field $\mathbf{E}^+ \propto (1,0,0) \equiv \mathbf{e}^+$ and $\mathbf{E}^- \propto (0,1,0) \equiv \mathbf{e}^-$ are eigenmodes of $\hat{\sigma}$ with the eigenvalues $\sigma = \pm 1$, whereas the third eigenvector of $\hat{\sigma}$ corresponds to the $z$-polarized field, $\mathbf{E}_z \propto (0,0,1)$, with the eigenvalue $\sigma = 0$. There are some fundamental mathematical difficulties in using canonical operators $\hat{\mathbf{L}}$ and $\hat{\mathbf{S}}$, Eq. (3), for generic nonparaxial fields [29,35,36], but they do not affect the OAM and SAM expectation values which will be calculated for different optical systems below.

## 3. High-NA focusing

Let us consider the Debye–Wolf theory of focusing with a spherical lens [37,38], Fig. 1. The incident field $\mathbf{E}_0(\mathbf{r})$ is paraxial, and one can neglect its $z$-component: $\mathbf{E}_0 \simeq (E_0^+, E_0^-, 0)^T$. Entering partial rays with the wave vector $\mathbf{k}_0 \simeq k\mathbf{u}_z$ are marked by coordinates at the entrance pupil, which can be expressed via spherical angles $(\tilde{\theta}, \tilde{\phi})$ defined with respect to the origin in the focal point: $x = f \sin\tilde{\theta}\cos\tilde{\phi}$, $y = f\sin\tilde{\theta}\sin\tilde{\phi}$, and $z = f\cos\tilde{\theta}$ ($f$ is the focal distance of the lens). After refraction, the partial rays converge at the focal point and have the nonparaxial **k**-vectors characterized by spherical coordinates $(\theta, \phi) = (\pi - \tilde{\theta}, \pi + \tilde{\phi})$ in the momentum space: $k_x = k\sin\theta\cos\phi$, $k_y = k\sin\theta\sin\phi$, and $k_z = k\cos\theta$ (see Fig. 1) [38]. The rays are distributed in the range $\{\theta \in (0, \theta_c), \phi \in (0, 2\pi)\}$, where $\theta_c$ is the aperture angle of the lens, and carry the electric fields $\tilde{\mathbf{E}}(\mathbf{k})$. Thus,

$$k_x = -k\frac{x}{f}, \quad k_y = -k\frac{y}{f}, \quad k_z = -k\frac{z}{f}, \tag{4}$$

and the lens performs a sort of *Fourier transform* translating the initial real-space distribution $\mathbf{E}_0(\mathbf{r})$ into the momentum distribution $\tilde{\mathbf{E}}(\mathbf{k})$ in the image space. In doing so, the angles $(\theta, \phi)$ serve both as real-space coordinates for the incident field $\mathbf{E}_0$ and momentum-space coordinates for the refracted field $\tilde{\mathbf{E}}$.

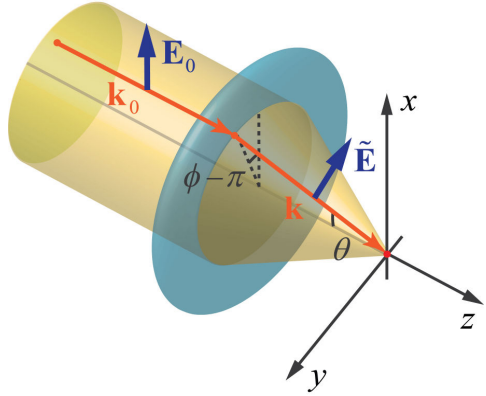

Fig. 1. Focusing of light by a spherical high-NA lens. The incident paraxial field $\mathbf{E}_0(x, y)$ is refracted in the meridional plane and transformed to a spectrum of plane waves (rays) $\tilde{\mathbf{E}}(\theta, \phi)$ with the $\mathbf{k}$-vectors distributed on the sphere in $\mathbf{k}$-space: $\theta \in (0, \theta_c)$, $\phi \in (0, 2\pi)$. The electric field vectors are parallel-transported along each partial ray with the local helicity being conserved.

The Debye–Wolf theory assumes that partial waves do *not* change their polarization state in the local basis attached to the ray, and the electric fields experience pure meridional rotations by the angle $\theta$ together with their **k**-vectors. This is an *adiabatic approximation* which neglects the polarization dependence of the refraction coefficients, cf. [39]. As a result, the focused field spectrum $\tilde{\mathbf{E}}(\mathbf{k})$ can be written using purely geometrical rotational transformation [16,40] $\hat{U}(\theta, \phi) = \hat{V}^\dagger \hat{R}_z(-\phi) \hat{R}_y(-\theta) \hat{R}_z(\phi) \hat{V}$, where $\hat{R}_a(\alpha) = \exp(i\alpha \hat{S}_a)$, $a = x, y, z$, is the matrix of rotation about the $a$-axis by the angle $\alpha$ [$(\hat{S}_a)_{ij} = -i\varepsilon_{aij}$ are the SO(3) generators of rotations, i.e., spin-1 matrices]. Explicitly, the transformation of the field in the circular basis takes the form [16,41]:

$$\tilde{\mathbf{E}} \propto \sqrt{\cos\theta}\, \hat{U}(\theta, \phi) \mathbf{E}_0, \quad \hat{U}(\theta, \phi) = \begin{pmatrix} a & -be^{-2i\phi} & \sqrt{2ab}\,e^{-i\phi} \\ -be^{2i\phi} & a & \sqrt{2ab}\,e^{i\phi} \\ -\sqrt{2ab}\,e^{i\phi} & -\sqrt{2ab}\,e^{-i\phi} & a-b \end{pmatrix}, \tag{5}$$

where $a = \cos^2(\theta/2)$, $b = \sin^2(\theta/2)$, and $\sqrt{\cos\theta}$ is the apodization factor that ensures the conservation of the energy flow [38]. Remarkably, the same unitary transformation $\hat{U}(\theta, \phi)$ describes transition from the global circular basis (1) to the local *helicity basis* attached to the partial wave vector for a generic nonparaxial field [29]. This reflects the fact that the refracted field does not change its polarization state in the helicity basis (labelled by subscript "*H*"):

$$\tilde{\mathbf{E}}_H = \hat{U}^\dagger \tilde{\mathbf{E}} \propto \mathbf{E}_0 \simeq \mathbf{E}_{0H} \,. \tag{6}$$

It follows from here that the high-NA focusing of a paraxial circularly-polarized field generates a nonparaxial *pure helicity state* of light. Recently, some of us considered nonparaxial vector Bessel beams with well-defined helicity [29], which could be generated, e.g., via focusing by an axicon lens with fixed $\theta = \theta_0$. The focusing with a spherical lens differs only by a smooth $\theta$-distribution of the produced plane-wave spectrum.

The three successive rotations, $\hat{R}_z(\phi)$, $\hat{R}_y(-\theta)$, and $\hat{R}_z(-\phi)$, in the transformation $\hat{U}$ indicate, respectively: (i) azimuthal rotation of the coordinate frame superimposing the $(x,z)$-plane with the local meridional plane, (ii) refraction of the field on the angle $\theta$ therein, and (iii) the reverse azimuthal rotation compensating the first one. Because of the noncommutativity of the rotations, this transformation is accompanied by a generation of geometrical phases which appear in the form of vortices in the off-diagonal elements of Eq. (5). These elements describe effective transitions between different AM modes. For instance, if the incident wave is $\sigma$-circularly polarized, $\mathbf{E}_0^\sigma \propto \mathbf{e}^\sigma$, i.e., has only the $E^\sigma$ component, the refracted wave is $\tilde{\mathbf{E}}^\sigma \propto \hat{U}\mathbf{E}_0^\sigma$, i.e., $\tilde{\mathbf{E}}^+ \propto \left(a, -be^{2i\phi}, -\sqrt{2ab}\,e^{i\phi}\right)^T$ and $\tilde{\mathbf{E}}^- \propto \left(-be^{-2i\phi}, a, -\sqrt{2ab}\,e^{-i\phi}\right)^T$. This indicates the generation of the $\tilde{E}_z$ component with the charge-$\sigma$ vortex $e^{i\sigma\phi}$ and the oppositely-polarized component $\tilde{E}^{-\sigma}$ with the charge-$2\sigma$ vortex $e^{2i\sigma\phi}$. We emphasize once again that the actual helicities of partial waves remain unchanged, Eq. (6), and these components appear because of the observation of the redirected rays in the same laboratory frame. Nonetheless, the azimuthal phases signify real generation of the OAM in the laboratory frame, i.e., the *spin-to-orbital AM conversion*. Let the incident wave be a paraxial vortex beam with circular polarization $\sigma$ and vortex charge $\ell$, and let us denote the AM state of light with respect to the $z$-axis by the OAM and SAM quantum numbers: $|\ell, \sigma\rangle$. Then, the polarization transformation (5) can be symbolically written as

$$|\ell, \sigma\rangle \to \sqrt{\cos\theta}\left[a|\ell,\sigma\rangle - b|\ell+2\sigma, -\sigma\rangle - \sqrt{2ab}\,|\ell+\sigma, 0\rangle\right]. \tag{7}$$

Equation (7) exhibits the conservation of the total AM quantum number in each term: $\sigma + \ell = \text{const}$, where the last term $|\ell+\sigma, 0\rangle$ corresponds to the longitudinal field $\tilde{E}_z$ which carries no SAM. Both the transverse field with the opposite polarization (*b*-term) and the longitudinal field ($\sqrt{2ab}$-term) contribute to the AM conversion. However, in the paraxial approximation, $\theta \ll 1$, one has $b \simeq \theta^2/4$ and $\sqrt{2ab} \simeq \theta/\sqrt{2}$, so that the main contribution is due to the $z$-component of the field [11].

First, we calculate the local (angle-resolved) *OAM and SAM densities*, $l_z$ and $s_z$, without integration over the $(\theta, \phi)$-distribution of the field. Using operators (3) in the circular basis, one can write

$$l_z(\theta, \phi) = -i\frac{\tilde{\mathbf{E}}^* \cdot \partial_\phi \tilde{\mathbf{E}}}{\tilde{\mathbf{E}}^* \cdot \tilde{\mathbf{E}}} \,, \quad s_z(\theta, \phi) = \frac{\tilde{\mathbf{E}}^* \cdot \hat{\sigma}\, \tilde{\mathbf{E}}}{\tilde{\mathbf{E}}^* \cdot \tilde{\mathbf{E}}} \,. \tag{8}$$

The electric field of the incident circularly-polarized vortex beam can be written as

$$\mathbf{E}_{0\ell}^\sigma \simeq \mathbf{e}^\sigma E_\ell(\theta, \phi), \quad E_\ell = F_{|\ell|}(\theta) e^{i\ell\phi}, \tag{9}$$

and $F_{|\ell|}(\theta) \propto \sin^{|\ell|}\theta$ for the vortex beams with the waist much larger than the entrance pupil.

From Eq. (5) the refracted field is $\tilde{\mathbf{E}}_\ell^\sigma \propto \sqrt{\cos\theta}\,\hat{U}(\theta,\phi)\mathbf{e}^\sigma E_\ell(\theta,\phi)$, and Eqs. (8) brings about

$$l_z(\theta) = \ell + \sigma\,\Phi_B(\theta), \quad s_z(\theta) = \sigma\left[1 - \Phi_B(\theta)\right], \quad \Phi_B(\theta) = (1-\cos\theta) = 2b. \quad (10)$$

Here $\Phi_B(\theta) = 2\pi\,\Phi_B(\theta)$ stands for the spin-redirection Berry phase (see, e.g., [39]) associated with the azimuthal distribution of partial rays with a fixed polar angle $\theta$. Such distribution corresponds to the circular contour $\mathbb{C} = \{\theta = \text{const}, \phi \in (0, 2\pi)\}$ on the sphere of directions $S^2 = \{\mathbf{k}/k\}$ and the Berry phase gained after traversing this contour is $\Phi_B = \oint_\mathbb{C}(1-\cos\theta)d\phi = 2\pi(1-\cos\theta)$ [29]. Equations (10) characterize a $\theta$-dependent spin-to-orbital AM conversion which vanishes as $\Phi_B \approx \theta^2/2$ in the paraxial limit $\theta \ll 1$. This conversion is described by the Berry-phase term, whose appearance is explained in [29] in terms of phase matching of the geometrical-optics rays and quantization of caustics.

To calculate the OAM and SAM expectation values, $\langle L_z \rangle$ and $\langle S_z \rangle$ (throughout the paper we imply values per photon), one has to integrate both numerators and denominators of Eqs. (8) over the spherical angles $(\theta, \phi)$. Implying integration $\int d\Omega \equiv \int_0^{2\pi} d\phi \int_0^{\theta_c} d\theta \sin\theta$, the OAM and SAM expectation values are

$$\langle L_z \rangle = -i\frac{\int d\Omega\, \tilde{\mathbf{E}}^* \cdot \partial_\phi \tilde{\mathbf{E}}}{\int d\Omega\, \tilde{\mathbf{E}}^* \cdot \tilde{\mathbf{E}}}, \quad \langle S_z \rangle = \frac{\int d\Omega\, \tilde{\mathbf{E}}^* \cdot \hat{\sigma}\, \tilde{\mathbf{E}}}{\int d\Omega\, \tilde{\mathbf{E}}^* \cdot \tilde{\mathbf{E}}}. \quad (11)$$

For the field $\tilde{\mathbf{E}}_\ell^\sigma$, this results in (cf. [29,15])

$$\langle L_z \rangle = \ell + \sigma\,\bar{\Phi}_B, \quad \langle S_z \rangle = \sigma\left(1 - \bar{\Phi}_B\right), \quad \bar{\Phi}_B = \left(1 - \cos\bar{\theta}\right), \quad (12)$$

where the averaged polar angle represents a measure of the directional spread of the field and is defined as [43,44]:

$$\cos\bar{\theta} = \frac{c\langle P_z \rangle}{\langle W \rangle} = \frac{\int_0^{\theta_c} |F_{|\ell|}(\theta)|^2 \cos^2\theta \sin\theta\, d\theta}{\int_0^{\theta_c} |F_{|\ell|}(\theta)|^2 \cos\theta \sin\theta\, d\theta}. \quad (13)$$

Here $\langle P_z \rangle$ and $\langle W \rangle$ are the expectation values of the longitudinal momentum and energy of the focused field, which are based on the operators $\hat{p}_z = k\cos\theta$ and $\hat{w} = \omega$. Evidently, both angle-resolved densities and integral values of the OAM and SAM satisfy the conservation of the total AM per one photon:

$$l_z + s_z = \langle L_z \rangle + \langle S_z \rangle = \ell + \sigma. \quad (14)$$

Evaluation of Eq. (13) for $F_{|\ell|}(\theta) \propto (\sin\theta)^{|\ell|}$ (i.e., the vortex-core distribution) brings about

$$\cos\bar{\theta} = \frac{2(1+|\ell|)}{\sin^{(2+2|\ell|)}\theta_c}\left[\frac{2^{|\ell|}|\ell|!}{(2|\ell|+3)!!} - \frac{1}{3}\cos^3\theta_c\,{}_2F_1\left(\frac{3}{2}, -|\ell|, \frac{5}{2}; \cos^2\theta_c\right)\right], \quad (15)$$

where ${}_2F_1(\alpha, \beta, \gamma; z)$ is the Gauss hypergeometric function. In the simplest case of the incident plane wave with $\ell = 0$ and $F_0(\theta) = 1$, Eqs. (15) yield $\cos\bar{\theta} = 2(1-\cos^3\theta_c)/(3\sin^2\theta_c)$. Then, for small aperture angle, $\theta_c \ll 1$, $\cos\bar{\theta} \approx 1 - \theta_c^2/4$ and $\langle L_z \rangle \approx \sigma\,\theta_c^2/4$. For maximal aperture $\theta_c = \pi/2$, one has $\cos\bar{\theta} = 2/3$, $\langle L_z \rangle = \sigma/3$, and the efficiency of the spin-to-orbital AM conversion reaches the value of $1/3$. For higher values of $|\ell|$, the efficiency of the conversion increases as the field becomes concentrated at higher angles $\theta$ (see Fig. 8 below).

Let us compare Eqs. (12) and (13) with other calculations of the OAM and SAM of the tightly focused circularly-polarized light [9,13,15,28,29]. First, our results differ from calculations [9] based on approach of [27] with a nonconserved total AM: $\langle L_z \rangle + \langle S_z \rangle \neq \ell + \sigma$. At the same time, the post-paraxial estimation $\langle L_z \rangle \approx \sigma \theta_c^2 / 4$ is analogous to the multipole expansion of a strongly focused Gaussian beam [13]. Equations (12) and (13) are similar to calculations in [15] and [28], but differ from the final result in [15], apparently due to an arithmetic inaccuracy therein. Finally, Eq. (12) is entirely analogous to the OAM and SAM of nonparaxial Bessel beams with well-defined helicity [29] modified by averaging over the polar angles $\theta$.

So far, we used only the plane-wave spectrum of the focused field, $\tilde{\mathbf{E}}(\mathbf{k})$. The actual real-space electric field near the focal point is determined by the interference of the partial plane waves and is given by the Debye integral similar to Fourier transform [37,38]:

$$\mathbf{E}(\mathbf{r}) \propto \int d\Omega\, \tilde{\mathbf{E}}(\theta,\phi) e^{i\Phi(\theta,\phi,\mathbf{r})}, \quad \Phi = k(x\sin\theta\cos\phi + y\sin\theta\sin\phi + z\cos\theta). \quad (16)$$

Here $\Phi = k|\mathbf{r} - \mathbf{R}| - kf \simeq -k\mathbf{r}\cdot\mathbf{R}/f$ is the optical phase gained along the path from the refraction point $\mathbf{R} = -f(\sin\theta\cos\phi, \sin\theta\sin\phi, \cos\theta)$ to the observation point $\mathbf{r} = (x,y,z)$ ($|\mathbf{r}| \ll \sqrt{f/k}$ and the common phase $kf$ is subtracted). Calculating the Debye integral (16) and then the intensity, $I_\ell^\sigma = \mathbf{E}_\ell^{\sigma*} \cdot \mathbf{E}_\ell^\sigma$, for the incident paraxial beam (9) with the focusing transformation (5), one can derive (cf. [15,29]):

$$I_\ell^\sigma(\rho,z) \propto \left\langle a J_\ell(\xi) \right\rangle^2 + \left\langle b J_{\ell+2\sigma}(\xi) \right\rangle^2 + 2\left\langle \sqrt{ab}\, J_{\ell+\sigma}(\xi) \right\rangle^2. \quad (17)$$

Here we denoted $\xi = k\rho\sin\theta$ ($\rho$ is the radial cylindrical coordinate) and $\langle ... \rangle \equiv \int_0^{\theta_c} d\theta \sin\theta \sqrt{\cos\theta}\, F_{|\ell|}(\theta) e^{ikz\cos\theta}$. Note that for pure helicity states under consideration the magnetic-field intensity in real space coincides with the electric-field intensity, so that Eq. (17) can be regarded as the total intensity of the electromagnetic field.

It is seen that the intensity distribution of the focused field depends on both the OAM (vortex) and SAM (helicity) quantum numbers, $\ell$ and $\sigma$. It is invariant with respect to the transformation $(\ell,\sigma) \to (-\ell,-\sigma)$ but neither with respect to $(\ell,\sigma) \to (\ell,-\sigma)$ nor $(\ell,\sigma) \to (-\ell,\sigma)$, which reflects $\ell \cdot \sigma$ symmetry typical of spin-orbit interaction. For a scalar wave, the intensity (17) would be $I_\ell(\rho,z) \propto \left\langle J_\ell(\xi) \right\rangle^2$. Nonparaxial vector terms proportional to $b = \Phi_B/2$ describe spin-orbit coupling between the main mode of the order $\ell$ to the modes of the order $(\ell+\sigma)$ and $(\ell+2\sigma)$ which appear owing to the azimuthal geometric phases in the matrix $\hat{U}$, Eq. (5). Dependence of the intensity distribution (17) on $\ell$ and $\sigma$ is closely related to the spin-to-orbital AM conversion and the OAM expectation value $\langle L_z \rangle$, Eq. (12). Namely, generalizing the Bessel-beam results of [29] for the averaged polar angle $\bar{\theta}$, Eq. (13), the "mechanical" radius $R$ corresponding to the orbital angular momentum $\langle L_z \rangle$ and underlying the focal intensity (17) can be estimated as

$$R_\ell^\sigma \simeq \frac{\langle L_z \rangle}{k\sin\bar{\theta}} = \frac{\ell + \sigma(1-\cos\bar{\theta})}{k\sin\bar{\theta}}. \quad (18)$$

Indeed, the transverse momentum $\langle P_\perp \rangle = k\sin\bar\theta$ represents the azimuthal momentum in the focal spot (the mean radial momentum vanishes there), and using mechanical definition $\mathbf{L} = \mathbf{P}\times\mathbf{R}$ one can write $\langle L_z \rangle = \langle P_\perp \rangle R$. Comparison of the $\sigma$-dependent fine structure of radii (18) with the wave intensities (17) is given in Fig. 2. Similar polarization-dependent properties of the focal intensities have appeared in [10,11,14,15]. In particular, it follows from Eq. (17) that the intensity of the beams with antiparallel OAM and SAM, $\ell\sigma < 0$, does not vanish on the axis $\rho = 0$ for $\ell = 1, 2$. This was experimentally verified in [45,46]. In general, the size of the focal spot for a strongly focused beam has a fundamental lower bound that depends not only on the beam's directional spread but also on its OAM and SAM, such that beams with antiparallel OAM and SAM can achieve tighter focal spots than those for which these AM are parallel [44]. Note that asymmetric focusing with a truncated lens or distribution of the incident light brings about the $\ell$- and $\sigma$-dependent transverse shifts of the focal-spot intensity centroid, which is proportional to $\langle L_z \rangle$, i.e., *orbital* and *spin Hall effects* [29,30,47].

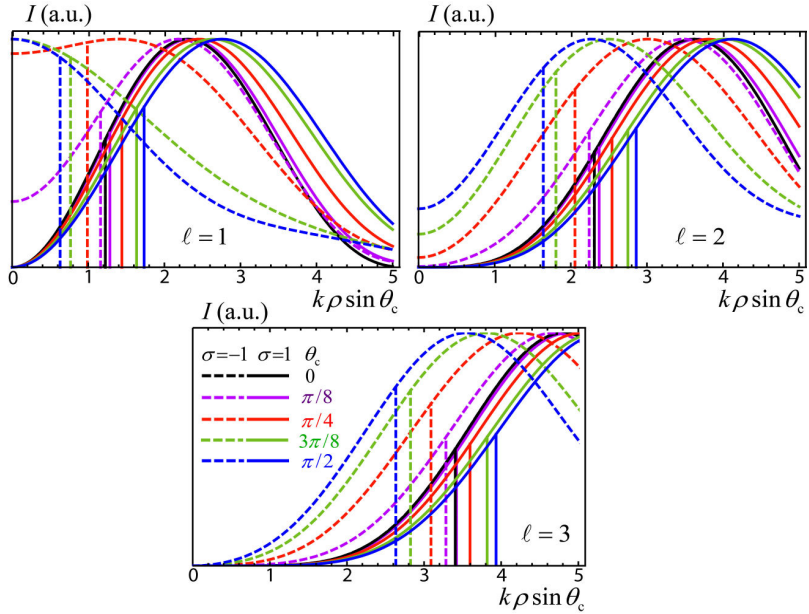

**Fig. 2.** Focal intensity distributions $I_\ell^\sigma(\rho,0)$, Eq. (17), in comparison with the radii $R_\ell^\sigma$, Eq. (18), (vertical lines) for $\ell = 1,2,3$, $\sigma = \pm 1$, and different values of the aperture angles $\theta_c$. It is seen that the structure of radii (18) underlies positions of the first maxima of intensity (17), cf. [29].

**4. Dipole scattering**

Scattering of paraxial light by a nano-particle located at the origin essentially represents the spherical redirection of partial plane waves, Fig. 3. We examine the simplest dipole approximation when the scattered spherical wave is generated by the dipole moment $\mathbf{p} \propto \mathbf{E}_0(\mathbf{0})$ proportional to the incident field $\mathbf{E}_0$ at the origin. Since higher-order paraxial beams (9) with $\ell \neq 0$ have zero field in the center, $\mathbf{E}_{0\ell}^\sigma(\mathbf{0}) \simeq \mathbf{0}$, below we consider an incident plane wave with circular polarization, $\mathbf{E}_0^\sigma(\mathbf{0}) \propto \mathbf{e}^\sigma$. The electric far field of the scattered wave is [48]

$$\tilde{\mathbf{E}}(\theta,\phi,r) \propto -\frac{\hat{\mathbf{r}} \times [\hat{\mathbf{r}} \times \mathbf{E}_0(\mathbf{0})]}{r}, \qquad (19)$$

where $\hat{\mathbf{r}} = \mathbf{r}/r$ is the unit radial vector with spherical coordinates $(\theta,\phi)$ of the observation point. Akin to focusing, the real-space spherical angles $(\theta,\phi)$ serve as the coordinates in momentum space for the scattered far field. In this manner, the $\hat{\mathbf{r}}$-vector plays the role of the $\mathbf{k}/k$-vector for scattered waves and transformation (19) is reminiscent of the vector transformation $\mathbf{k} \times (\mathbf{k} \times \tilde{\mathbf{E}})$ in free-space Maxwell equations $\mathbf{k} \times (\mathbf{k} \times \tilde{\mathbf{E}}) + k^2 \tilde{\mathbf{E}} = 0$. The double vector product $-\hat{\mathbf{r}} \times (\hat{\mathbf{r}} \times \mathbf{E}_0) = \mathbf{E}_0 - \hat{\mathbf{r}}(\hat{\mathbf{r}} \cdot \mathbf{E}_0)$ represents a *spherical projection* of $\mathbf{E}_0(\mathbf{0})$ onto the $\hat{\mathbf{r}}$-sphere and can be written, in the laboratory circular basis (1), as the action of a nondiagonal operator $\hat{\Pi}(\theta,\phi) = \hat{U}(\theta,\phi) \hat{\mathcal{P}}_z \hat{U}^\dagger(\theta,\phi)$ [16]:

$$\tilde{\mathbf{E}}(\theta,\phi) \propto \frac{1}{r} \hat{\Pi}(\theta,\phi) \mathbf{E}_0(\mathbf{0}), \quad \hat{\Pi} = \frac{1}{2} \begin{pmatrix} 1+a_1 & -b_1 e^{-2i\phi} & -\sqrt{2a_1 b_1} e^{-i\phi} \\ -b_1 e^{2i\phi} & 1+a_1 & -\sqrt{2a_1 b_1} e^{i\phi} \\ -\sqrt{2a_1 b_1} e^{i\phi} & -\sqrt{2a_1 b_1} e^{-i\phi} & 2b_1^2 \end{pmatrix}. \qquad (20)$$

Here $\hat{\mathcal{P}}_z = \mathrm{diag}(1,1,0)$ is the projector onto the orthogonal plane, $a_1 = \cos^2\theta$, and $b_1 = \sin^2\theta$.

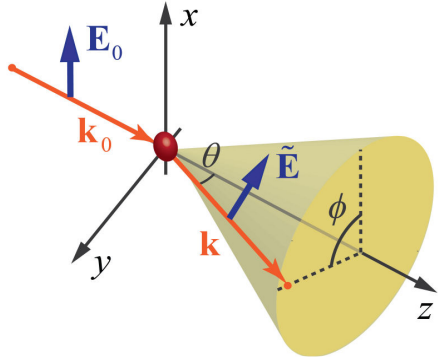

**Fig. 3.** Akin to focusing, Fig. 1, dipole scattering of the incident paraxial field $\mathbf{E}_0(x,y)$ by a small spherical particle transforms it into a spectrum of plane waves $\tilde{\mathbf{E}}(\theta,\phi)$ with the spherically-distributed $\mathbf{k}$-vectors.

Projection (20) is a nonunitary transformation, and it does not preserve the local helicity of the scattered field. Indeed, in the helicity basis attached to the sphere of partial $\mathbf{k}$-vectors, the scattered field is [cf. Eq. (6)]:

$$\tilde{\mathbf{E}}_H = \hat{U}^\dagger \tilde{\mathbf{E}} \propto \hat{\mathcal{P}}_z \hat{U}^\dagger \mathbf{E}_0 \simeq \hat{\mathcal{P}}_z \hat{U}^\dagger \mathbf{E}_{0H}. \qquad (21)$$

Other than that, Eq. (20) demonstrates features that are quite similar to those of the focusing transformation (5). In particular, the off-diagonal geometric-phase elements of the matrix $\hat{\Pi}$ produces *spin-to-orbital AM conversion* which can be symbolically written similarly to Eq. (7):

$$|0,\sigma\rangle \to \frac{1}{2}\left[(1+a_1)|0,\sigma\rangle - b_1|2\sigma,-\sigma\rangle - \sqrt{2a_1 b_1}|\sigma,0\rangle\right]. \qquad (22)$$

The AM conversion upon scattering of light on various objects was analyzed in several papers [16-24]. While in the case of focusing the polarization is not changed in the helicity basis, the transformation (21) of the scattered field can be written as the following helicity transition:

$$|\sigma\rangle_H \to a|\sigma\rangle_H - be^{2i\sigma\phi}|-\sigma\rangle_H. \qquad (23)$$

In the linear approximation in $\theta$ (for small scattering angles $\theta \ll 1$), the helicity (23) is conserved [19,20], and the scattering transformation (20) becomes similar to the focusing one, Eq. (5).

Assuming an incident plane wave with circular polarization, $\mathbf{E}_0^\sigma(\mathbf{0}) \propto \mathbf{e}^\sigma$, we determine the scattered field (20) and calculate the OAM and SAM angle-resolved densities using Eqs. (8):

$$l_z = \sigma \frac{1-\cos^2\theta}{1+\cos^2\theta}, \quad s_z = \sigma \frac{2\cos^2\theta}{1+\cos^2\theta}. \qquad (24)$$

Identical results were obtained for the field radiated by a rotating dipole [49], and similar results [but without $(1+\cos^2\theta)^{-1}$ factor] have appeared for the Rayleigh scattering [18]. Equations (24) yield $l_z \approx \theta^2\sigma$ for the paraxial angles $\theta \ll 1$ and the total conversion $l_z = \sigma$ at $\theta = \pi/2$ (see Fig. 8 below). The integral OAM and SAM values are determined via Eqs. (11) with the spherical integration $\int d\Omega \equiv \int_0^{2\pi} d\phi \int_0^\pi d\theta \sin\theta$, which results in

$$\langle L_z \rangle = \sigma \frac{\int_0^\pi (1-\cos^2\theta)\sin\theta\, d\theta}{\int_0^\pi (1+\cos^2\theta)\sin\theta\, d\theta} = \frac{1}{2}\sigma, \quad \langle S_z \rangle = \sigma \frac{2\int_0^\pi \cos^2\theta \sin\theta\, d\theta}{\int_0^\pi (1+\cos^2\theta)\sin\theta\, d\theta} = \frac{1}{2}\sigma. \qquad (25)$$

Thus, the spin-to-orbital AM conversion has the efficiency $1/2$ [18,49] (see Fig. 8 below). Equations (24) and (25) demonstrate conservation of the local and integral total AM per photon, Eq. (14).

It is worth remarking that the converted part of the AM in Eq. (24) can be expressed as

$$l_z = \sigma P(\theta), \quad s_z = \sigma[1-P(\theta)], \quad P(\theta) = \frac{2\mathcal{I}^{-\sigma}}{\mathcal{I}^\sigma + \mathcal{I}^{-\sigma}} = \frac{\sin^2\theta}{1+\cos^2\theta}, \qquad (24a)$$

where $\mathcal{I}^\sigma = a^2$ and $\mathcal{I}^{-\sigma} = b^2$ stand for local intensities of the helicity components, Eq. (23). Here the quantity $P(\theta)$ coincides with the *degree of polarization* of the scattered unpolarized light [48]. Thus, the changes in the degree of polarization upon scattering are also connected with the above-considered geometric transformations of the wave field. In particular, the depolarization of multiply scattered polarized light and typical four-fold polarization patterns of the backscattered light are intimately related to the spin-to-orbital AM conversion [19,20]. In the weak-scattering approximation of small scattering angles ($\theta \ll 1$ in a single scattering), these depolarization effects can be explained via Berry-phase accumulation along the partial scattering paths [19,20,50,51]. This establishes a geometric-phase link between the AM conversions in focusing and scattering processes. For strong single scattering event ($\theta \sim 1$), the Berry-phase (adiabatic) approximation is not applicable because the geometric transformation (20) represents a projection rather than parallel-transport rotation (5) of the field.

## 5. Imaging of nanoparticles

Strongly focused or scattered fields are essentially nonparaxial, which is the main source of the spin-orbit phenomena. Accordingly, the AM conversion can be detected via various near-field methods: e.g., tracing motion of testing particles [11,12] or using near-field probes

[14,21,24,47]. It should be noted that the use of testing particles is somewhat ambiguous as they can undergo orbital motion due to *both* orbital and spin energy flows, i.e., OAM and SAM [52,53]. At the same time, traditional paraxial-optics detectors are unable to measure adequately the spin-orbit phenomena in nonparaxial fields with strong longitudinal field component. It turns out, however, that a standard imaging system successfully resolves this dilemma by transfering the spin-orbit coupling into a paraxial far-field, where it can be easily detected. There were several experimental observations [16,31-34,54] of the spin-orbit interactions of light using imaging scheme: (i) focusing of the incident paraxial light with a high-NA lens; (ii) scattering by a small specimen; (iii) collection of the scattered light by another high-NA lens transforming it to the outgoing paraxial light, see Fig. 4. In most cases the observed effects were ascribed either to focusing or to scattering process, although all the three elements of the imaging system contributed to the effect. The 'lens-scatterer-lens' system (Fig. 4) represents the basis for optical microscopy and it is important to describe the spin-orbit effects in such imaging system taking into account all its elements [16,17].

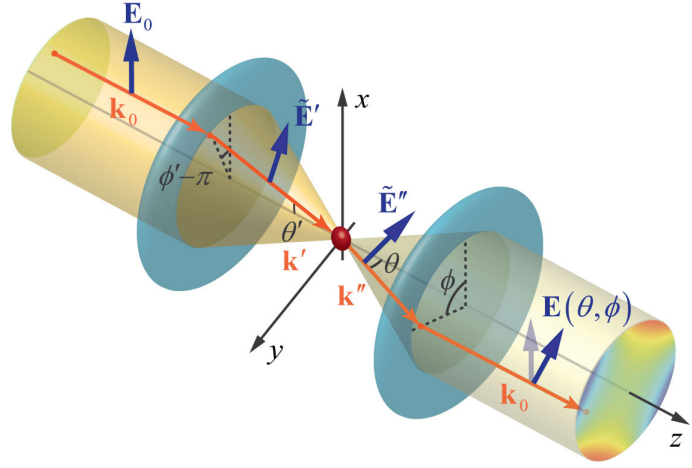

**Fig. 4.** Scheme of the 'lens-scatterer-lens' imaging system. First, the incident paraxial light $\mathbf{E}_0$ is focused by a high-NA lens, then a small specimen in the focus scatters the nonparaxial focused field, and finally the scattered light is collected by the second high-NA lens. The output paraxial field $\mathbf{E}(\theta,\phi)$ has a space-variant polarization distribution and bears information about the spin-orbit coupling inside the system. This offers an efficient tool to retrieve fine subwavelength information about the specimen.

Since both the input and output fields in the imaging system are paraxial, the transformation of the field by the system can be described by an effective *Jones matrix* which gives the angle-resolved polarization state of the output field. Using successive applications of the geometric transformations of the first lens, Eq. (5), together with the Debye integral (16), dipole-scattering transformation (20), and the inverse transformation (5) of the collector lens, one can obtain the 3D transformation operator of the system [16], $\mathbf{E}(\theta,\phi) \propto \hat{T}(\theta,\phi)\mathbf{E}_0(\theta',\phi')$:

$$\hat{T} = \frac{1}{\sqrt{\cos\theta}}\hat{\mathcal{P}}_z\hat{U}^\dagger(\theta,\phi)e^{i\Phi(\theta,\phi,\mathbf{r}_s)}\int d\Omega' \sqrt{\cos\theta'}\,\hat{U}(\theta',\phi')e^{i\Phi'(\theta',\phi',\mathbf{r}_s)}. \qquad (26)$$

Here $\int d\Omega' \equiv \int_0^{2\pi}d\phi' \int_0^{\theta_c}d\theta'\sin\theta'$, the angles $(\theta',\phi')$ and $(\theta,\phi)$ mark the partial rays at the entrance and exit pupils, respectively (Fig. 4), $\mathbf{r}_s = (x_s, y_s, z_s)$ is the position of the scattering particle near the common focus of the two lenses, and the phases are

$$\Phi' \simeq -k\,\mathbf{r}_s \cdot \mathbf{R}'/f = k\left(x_s \sin\theta' \cos\phi' + y_s \sin\theta' \sin\phi' + z_s \cos\theta'\right),$$
$$\Phi \simeq -k\,\mathbf{r}_s \cdot \mathbf{R}/f = -k\left(x_s \sin\theta \cos\phi + y_s \sin\theta \sin\phi + z_s \cos\theta\right). \quad (27)$$

Here $\mathbf{R}' = -f\left(\sin\theta'\cos\phi', \sin\theta'\sin\phi', \cos\theta'\right)$ and $\mathbf{R} = f\left(\sin\theta\cos\phi, \sin\theta\sin\phi, \cos\theta\right)$ are the refraction points at the focusing and collecting lenses, respectively. The projector $\hat{\mathcal{P}}_z$ in Eq. (26) ensures transversality of the output field, and the upper left $2\times 2$ sector of the operator (26), $\hat{T}_\perp(\theta,\phi)$, provides the effective space-variant Jones matrix connecting the transverse components of the input and output fields in the basis of circular polarizations. In what follows we denote the transverse components of the vectors and matrices by the subscript "$\perp$": $\mathbf{E}_\perp = \left(E^+, E^-\right)^T$, etc. Note that even if the incident light represents a homogeneous plane wave $\mathbf{E}_0$, the output field will have nonuniform polarization $\mathbf{E}(\theta,\phi)$ bearing information about the spin-orbit coupling in the system and specimen properties.

First, let us consider the symmetric case when the scatterer is located precisely in the focus: $\mathbf{r}_s = \mathbf{0}$. If the incident field is a homogeneous plane wave $\mathbf{E}_0$ (which implies $\ell = 0$), one can evaluate the integral (26) analytically. The resulting Jones operator is [16]:

$$\hat{T}_\perp^{(0)} = \frac{A(\theta_c)}{\sqrt{\cos\theta}} \begin{pmatrix} a & -be^{-2i\phi} \\ -be^{2i\phi} & a \end{pmatrix}, \quad (28)$$

where the aperture-dependent coefficient is $A(\theta_c) = \frac{2\pi}{15}\left[8 - (\cos\theta_c)^{3/2}(5 + 3\cos\theta_c)\right]$ and $A \approx \pi\theta_c^2$ at $\theta_c \ll 1$. The Jones matrix (28) is proportional to the transverse part of $\hat{U}$, Eq. (5), and it describes the *spin-to-orbital AM conversion* between two *paraxial* states of light:

$$|0,\sigma\rangle \to a|0,\sigma\rangle - b|2\sigma, -\sigma\rangle. \quad (29)$$

Comparing this equation with Eq. (23), one can see that, seemingly, the imaging scheme transfers the scattering-induced AM conversion which appears in the local helicity basis to the paraxial field and laboratory circular basis. AM conversion (29) strongly resembles paraxial spin-to-orbital convertors based on locally-anisotropic axially symmetric structures [3-7,25,26]. In our case, the $\phi$-dependent helical phases arise from purely geometrical 3D transformations of the field inside the imaging system, which demonstrate effective birefringence of the radially (TM) and azimuthally (TE) polarized modes. Assuming $\sigma$-circularly polarized incident wave, $\mathbf{E}_0^\sigma \propto \mathbf{e}^\sigma$, the output field is $\mathbf{E}_\perp^\sigma \propto \hat{T}_\perp^{(0)} \mathbf{e}_\perp^\sigma$, i.e., $\mathbf{E}_\perp^+ \propto \left(a, -be^{2i\phi}\right)^T/\sqrt{\cos\theta}$ and $\mathbf{E}_\perp^- \propto \left(-be^{-2i\phi}, a\right)^T/\sqrt{\cos\theta}$. Using Eqs. (8), we determine the local OAM and SAM densities at the exit pupil:

$$l_z = \sigma\left(1 - \frac{2\cos\theta}{1+\cos^2\theta}\right), \quad s_z = \sigma\frac{2\cos\theta}{1+\cos^2\theta}. \quad (30)$$

From Eqs. (11), implying integration over the exit pupil, $\int dx\, dy \propto \int d\Omega_\perp \equiv \int_0^{2\pi} d\phi \int_0^{\theta_c} d\theta \sin\theta \cos\theta$, one can obtain the integral OAM and SAM values:

$$\langle L_z \rangle = \sigma\frac{1 - 3\cos\theta_c + 3\cos^2\theta_c - \cos^3\theta_c}{4 - 3\cos\theta_c - \cos^3\theta_c}, \quad \langle S_z \rangle = \sigma\frac{3\sin^2\theta_c}{4 - 3\cos\theta_c - \cos^3\theta_c}, \quad (31)$$

which satisfies the total AM conservation (14). The efficiency of the spin-to-orbital AM conversion is tiny in the imaging system with small aperture: $\langle L_z \rangle \approx \sigma\theta_c^4/24$ at $\theta_c \ll 1$, and it reaches the value $\langle L_z \rangle = \sigma/4$ at $\theta_c = \pi/2$.

Since the output field $\mathbf{E}(\theta,\phi)$ is paraxial, its polarization properties can be characterized by space-variant Stokes parameters. The normalized Stokes vector $\vec{\mathfrak{S}} = (\mathfrak{S}_1, \mathfrak{S}_2, \mathfrak{S}_3)$ can be calculated using the Pauli matrices $\hat{\vec{\sigma}} = (\hat{\sigma}_1, \hat{\sigma}_2, \hat{\sigma}_3)$ [55]:

$$\vec{\mathfrak{S}}(\theta,\phi) = \frac{\tilde{\mathbf{E}}^* \cdot \hat{\vec{\sigma}} \tilde{\mathbf{E}}}{\tilde{\mathbf{E}}^* \cdot \tilde{\mathbf{E}}}, \qquad (32)$$

Note that the spin density $s_z$, Eq. (30), coincides with the local degree of circular polarization, i.e., the third component of the Stokes vector: $\mathfrak{S}_3 = s_z$. Indeed, for paraxial fields it is determined by the transverse part of the operator $\hat{\sigma}$: $\hat{\sigma}_\perp = \mathrm{diag}(1,-1) \equiv \hat{\sigma}_3$. The complete description of the polarization contains more information than the spin AM of light [56]. In particular, the first two components of the Stokes vector (32) can evidence conversion to the OAM, Eq. (29). Indeed, calculating Eq. (32) for the incident circularly-polarized plane wave, we obtain

$$\vec{\mathfrak{S}}(\theta,\phi) = \frac{1}{a^2+b^2}\left[-2ab\cos 2\phi, -2ab\sin 2\phi, \sigma(a^2-b^2)\right]. \qquad (33)$$

This distribution of the Stokes parameters is shown in Fig. 5. Parameters $\mathfrak{S}_1$ and $\mathfrak{S}_2$ demonstrate typical four-fold patters which signify generation of the oppositely polarized component $|-\sigma\rangle$ with the optical vortex $e^{2i\sigma\phi}$ [16,19,20,33,50].

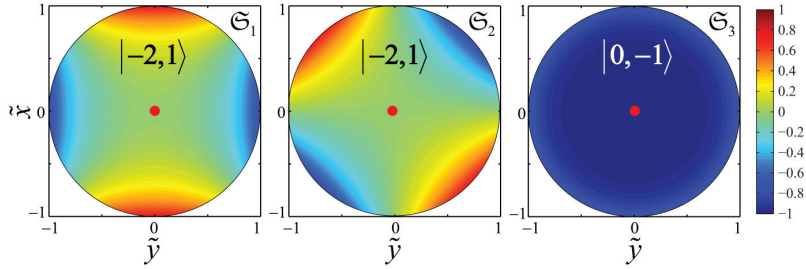

**Fig. 5.** Distributions of the Stokes parameters $\vec{\mathfrak{S}} = (\mathfrak{S}_1, \mathfrak{S}_2, \mathfrak{S}_3)$, Eqs. (32) and (33), in the exit pupil of the optical microscope (Fig. 4) in the case of the on-axis location of the specimen, $\mathbf{r}_s = \mathbf{0}$, and left-hand circular polarization of the incident light, $\sigma = -1$. The normalized coordinates $\tilde{x} = x/(f\sin\theta_c)$ and $\tilde{y} = y/(f\sin\theta_c)$ are used. The four-fold patterns in the $\mathfrak{S}_1$ and $\mathfrak{S}_2$ distributions are the signature of the generation of the right-hand polarized component with optical vortex $e^{-2i\phi}$, i.e., the spin-to-orbital AM conversion (29). The aperture angle is $\theta_c = 3\pi/8$.

The cylindrical symmetry is broken in the imaging system if the specimen is transversely shifted in the focal plane. In this case, the system is not rotationally-invariant about the $z$-axis, and the total AM is no longer conserved. This results in the AM-dependent orthogonal shift of the center of gravity of the output field, i.e., the Hall effect of light. Considering a small subwavelength displacement of the specimen in the $(x,y)$ plane, $\mathbf{r}_s = (x_s, y_s, 0)$, $k|\mathbf{r}_s| \ll 1$, the effective Jones matrix can be evaluated analytically from Eq. (26) as a correction to Eq. (28) [16]: $\hat{T}_\perp \simeq \hat{T}_\perp^{(0)} + \hat{T}_\perp^{(1)}$,

$$\hat{T}_\perp^{(1)} = -ik\frac{B(\theta_c)\sin\theta}{2\sqrt{\cos\theta}}\begin{pmatrix} \rho_s e^{-i\phi} & \rho_s^* e^{-i\phi} \\ \rho_s e^{i\phi} & \rho_s^* e^{i\phi} \end{pmatrix}. \qquad (34)$$

Here we introduced $\rho_s = x_s + iy_s$ and the aperture-dependent coefficient is $B(\theta_c) = \frac{\pi}{21}\left[8 - (\cos\theta_c)^{3/2}(11 - 3\cos 2\theta_c)\right]$, with $B \approx \pi\theta_c^4/4$ at $\theta_c \ll 1$. The Jones matrix (34) reveals coupling between the position of the scatterer and polarization of light, which results in the *spin Hall effect*. Taking the incident circularly polarized plane wave $\mathbf{E}_0^\sigma \propto \mathbf{e}^\sigma$, we assume $\rho_s = x_s$, determine the output field $\mathbf{E}_\perp^\sigma \propto \hat{T}_\perp \mathbf{e}_\perp^\sigma$, and calculate the transverse position of centre of gravity of the field. It is determined as

$$\langle \mathbf{R}_\perp \rangle = \frac{\int d\Omega_\perp \mathbf{R}_\perp \left(\mathbf{E}^{\sigma*} \cdot \mathbf{E}^\sigma\right)}{\int d\Omega_\perp \mathbf{E}^{\sigma*} \cdot \mathbf{E}^\sigma}, \quad (35)$$

where $\mathbf{R}_\perp \equiv (X, Y) = f(\sin\theta\cos\phi, \sin\theta\sin\phi)$. Substituting the output field calculated through the Jones matrix (28) and (34) into Eq. (35), we obtain

$$\langle X \rangle = 0, \quad \langle Y \rangle = -\sigma f k x_s \frac{3B\sin^4\theta_c}{2A(4 - 3\cos\theta_c - \cos^3\theta_c)}. \quad (36)$$

The $\sigma$-dependent shift, orthogonal to the displacement of the specimen, represents the spin Hall effect of light. It is also accompanied by a nonzero mean momentum $\langle P_x \rangle \neq 0$. The shift (36) behaves as $\langle Y \rangle \approx -\sigma f k x_s \theta_c^4/8$ at $\theta_c \ll 1$ and reaches value $\langle Y \rangle \simeq -0.13\sigma f k x_s$ at $\theta_c = \pi/2$. Thus, for high-NA systems and subwavelength displacements of the particle, $kx_s \sim 1$, the shift of the centre of gravity is of the order of a fraction of the focal length $f \gg \lambda$. In other words, high-NA optical microscope dramatically *magnifies* the spin Hall effect from the typical subwavelength scale to the scale of the exit pupil [16]. Figure 6 displays the intensity distributions $I = \mathbf{E}^{\sigma*} \cdot \mathbf{E}^\sigma$ and shifts of the field centroid at different positions $x_s$ of the scatterer. Similar $\ell$-dependent deformations of intensity in the imaging system with incident paraxial vortex beams, i.e., *orbital Hall effect*, were observed in [32]. Another striking manifestation of the spin Hall effect is a strong *separation* of local SAM densities in the incident *linearly* polarized light [16].

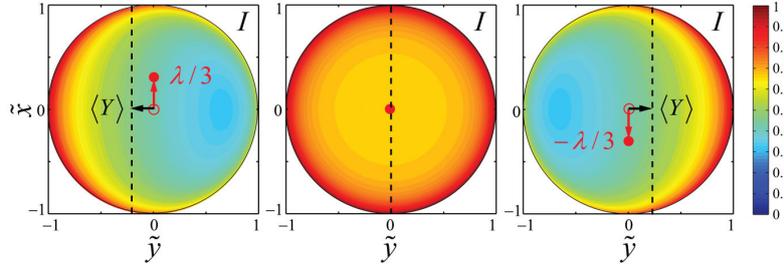

**Fig. 6.** Distributions of the output field intensity $I(\theta,\phi)$ in the imaging system Fig. 4 for the incident right-hand circularly polarized light ($\sigma = 1$) at different subwavelength displacements of the scattering particle: $x_s = 0, -\lambda/3, \lambda/3$. The transverse shift of the center of gravity of the field, $\langle Y \rangle \propto \sigma x_s$ signifies the spin Hall effect of light. Parameters are the same as in Fig. 5.

Alongside the Stokes polarimetry, which reveals conversion to different polarization and OAM modes, the state of the output light can be fully described via intensities and phases of the two polarization components following from the Jones matrices (28) and (34). Distributions of the intensities and phases of the right- and left-hand circularly polarized field components for the on-axis and off-axis scatterer are shown in Fig. 7. It is seen that the nonzero intensity and the vortices in the oppositely polarized (with respect to the incident light) component indicates the spin-to-orbital AM conversion (29). At the same time, the intensity redistribution and phase gradient in the main polarization component are responsible for the shifts (36), $\langle Y \rangle \neq 0$, $\langle P_x \rangle \neq 0$ for an off-axis scatterer.

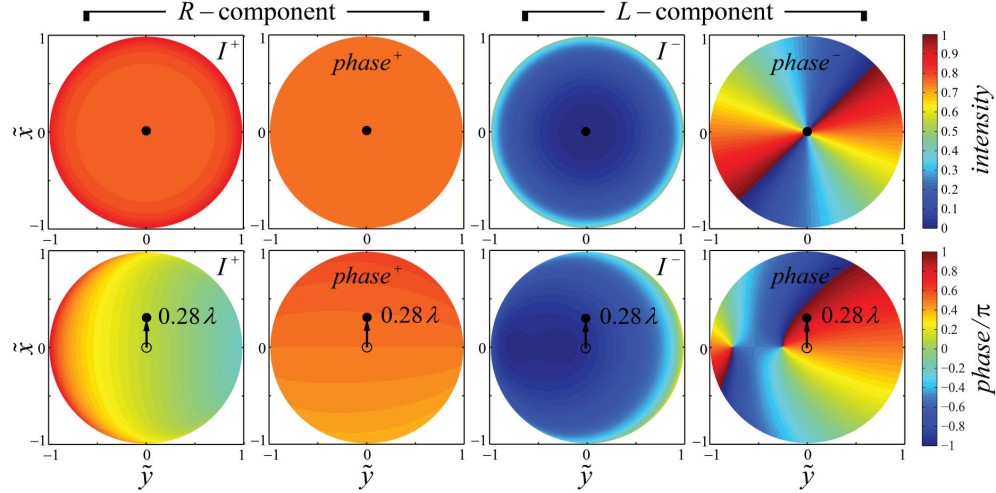

**Fig. 7.** Intensities and phases of the right- and left-hand circularly polarized components in the output field for the on-axis ($\mathbf{r}_s = \mathbf{0}$) and off-axis ($x_s = 0.28\lambda$) positions of the scatterer (Fig. 4). The incident field is right-hand circularly polarized ($\sigma = 1$), the aperture of the system corresponds to $\sin\theta_c = 0.922$. The charge-2 optical vortex and nonzero intensity in the left-hand polarized component is clearly seen for the on-axis particle. Displacement of the scatterer induces splitting of the charge-2 vortex into two charge-1 vortices (cf. [34]), strong deformation of the intensity of the right-hand component (responsible for the spin-Hall effect and $\langle Y \rangle \neq 0$), and smooth gradient of the phase in the right-hand component which yields $\langle P_x \rangle \neq 0$.

## 6. Conclusions

We have examined the spin-to-orbital AM conversion in basic optical systems involving nonparaxial fields. The AM conversion originates from geometric transformations of the wave field: parallel-transport rotations in the case of focusing and spherical projections in the case of scattering. Despite the fact that problems of spin-to-orbital conversion were considered in recent years both in focusing and scattering systems, a number of controversies and the use of dissimilar approaches did not allow one to obtain solid quantitative results and compare the efficiency of the conversion in different systems. Our approach unifies various treatments of the AM in different nonparaxial optical systems and is based on the geometric field transformations and canonical AM operators. As a result, one can compare the efficiency of the spin-to-orbital conversion in high-NA lens focusing, Rayleigh (dipole) scattering, and in far-field "lens-scatterer-lens" imaging system. Figure 8 shows both angle-resolved OAM and SAM densities and integral OAM and SAM values (per photon) as dependent on the aperture angle. For comparison we also plot there the OAM and SAM of nonparaxial Bessel beams calculated in [29]. One can see that for the plane incident wave ($\ell = 0$) the conversion efficiency decreases in a sequence: "Bessel beam" – "scattering" – "focusing" – "imaging". Curiously, the corresponding maximum efficiencies reached at the aperture angle $\theta_c = \pi/2$ are equal to 1, 1/2, 1/3, and 1/4. The total spin-to-orbital conversion for Bessel modes with $\theta_c = \pi/2$ was observed in a plasmonic experiment [14].

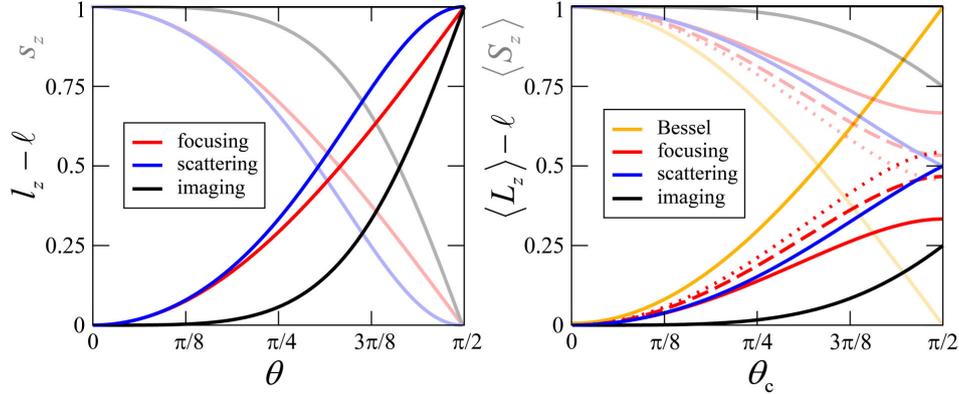

**Fig. 8.** Left: OAM angle-resolved density $l_z(\theta)$ converted from SAM for the cases of focusing (red) [Eq. (10)], scattering (blue) [Eq. (24)], and imaging (black) [Eq. (30)], with circularly polarized incident light, $\sigma = +1$, $\ell = 0$. Right: The integral OAM, $\langle L_z \rangle - \ell$, converted from SAM, vs. the aperture angle $\theta_c$ for focusing [Eqs. (12) and (13)], scattering [Eq. (25)], and imaging systems [Eq. (31)]. The solid, dashed, and dotted curves for focusing represent incident paraxial beams with $\ell = 0$, $\ell = 1$, and $\ell = 2$ ($\sigma = +1$ everywhere). For scattering, the dependence is obtained by formally replacing the upper limit of integration in Eqs. (25): $\pi \to \theta_c$; analytical results above are recovered at $\theta_c = \pi$ or $\pi/2$. The semitransparent curves indicate the corresponding quantities for the SAM, i.e., $s_z(\theta)$ and $\langle S_z \rangle$, satisfying the conservation law (14). For comparison, the right-hand panel also displays the OAM and SAM values for nonparaxial vector Bessel beams (yellow) [29], which achieve the total spin-to-orbital conversion at the aperture angle $\theta_c = \pi/2$, see [14].

Although the OAM and SAM are important dynamical characteristics of light, it is difficult to measure them directly in optical systems. In particular, orbital motion of small testing particles [11,12] cannot be used as a measure of the OAM because both the spin and orbital energy flows can be responsible for it [52,53]. However, the close connection between the intensity distributions (in particular the radius of the focal spot) and the OAM values, Eq. (18) and Fig. 2, enables one to quantify the AM conversion via spin-dependent intensity profiles. Also, as we have shown, the angle-resolved polarimetry in the far-field imaging systems enables one to reconstruct the field distribution and unambiguously characterize its AM features [16] (Section 5).

Importantly, universal mechanisms of the spin-to-orbital AM conversion manifest themselves not only in optical but also in plasmonic [14,22-24,47] and semiconductor [34] systems. Therefore, the presented unified geometric and operator description of the AM conversions provides a useful link between spin-orbit phenomena in nonparaxial wave fields of various nature.

**Acknowledgements**

This work was supported by the European Commission (Marie Curie Action), Science Foundation Ireland (Grant No. 07/IN.1/I906), and the Australian Research Council.